\documentclass[12pt]{iopart}

\usepackage{graphicx}
\usepackage[sorting=none,style=numeric,citestyle=numeric-comp]{biblatex}
\usepackage{amsmath}
\usepackage{graphicx}
\usepackage{caption}
\usepackage{subcaption}
\usepackage{multirow}
\usepackage{tikz}
\usepackage{amssymb}
\usepackage{multirow}
\usepackage{subcaption}

\begin{document}

\title[]{Bayesian Selection for Efficient MLIP Dataset Selection}

\author[]{Thomas Rocke$^1$\textsuperscript{\textdagger} and James R. Kermode$^1$}
\vspace{10pt}
\begin{indented}
\item[] 1. Warwick Centre for Predictive Modelling, School of Engineering, University of Warwick, Coventry CV4 7AL,
United Kingdom
\item[]
\item[] \textdagger {} Corresponding Author. Email: thomas.rocke@warwick.ac.uk
\item[] 
\item[] June 2025
\end{indented}

\begin{abstract}
The problem of constructing a dataset for MLIP development which gives the maximum quality in the minimum amount of compute time is complex, and can be approached in a number of ways. We introduce a ``Bayesian selection" approach for selecting from a candidate set of structures, and compare the effectiveness of this method against other common approaches in the task of constructing ideal datasets targeting Silicon surface energies. We show that the Bayesian selection method performs much better than Simple Random Sampling at this task (for example, the error on the (100) surface energy is 4.3x lower in the low data regime), and is competitive with a variety of existing selection methods, using ACE \cite{ACE} and MACE \cite{MACE} features.
\end{abstract}

%
%
\submitto{\MSMSE}
%
%
%

\section{Introduction}
Efficient dataset construction forms an important part of generating Machine Learning Interatomic Potentials (MLIPs) for new materials. The greatest contributor to the computational cost of adding structures to a dataset is almost always the evaluation of DFT and other QM methods to obtain energy and force data, rather than in the generation of atomic positions alone. 

Given a set of candidate structures, we can leverage sampling techniques to attempt to select a subset of structures which will maximally improve the quality of a MLIP model whilst requiring the least computational time possible. There are many different routes to this initial set of candidate structures, for example running Molecular Dynamics (MD) simulations with a potential, using random atom or cell displacements, random swaps of atoms, or generating structures using some continuum displacement law or experimentally known reconstruction. With the recent development of several ``foundation model" \cite{MP0, OMatEQV2, ORB, Mattersim, Gnome, Grace} potentials (models which attempt to describe all of materials chemistry), generating sets of candidate structures prior to performing DFT is increasingly accessible for more complex systems and properties.

Many approaches to the general problem of efficient dataset generation have already been proposed \cite{SubramanyamPerez,ACEHAL,Shapeev1,Novikov,Schaaf,Qi,Hodapp1,Hodapp2}. Subramanyam and Perez \cite{SubramanyamPerez} propose a scheme attempting to maximise the informational entropy of descriptor vectors contained in the dataset. Candidate structures are generated randomly, and the log determinant of the empirical covariance of descriptor vectors is used to define the informational entropy, assuming the descriptor vectors are independent and identically distributed (iid). 

The ACE HyperActive Learning approach (HAL/ACEHAL) \cite{ACEHAL} by van der Oord et. al. uses a committee of Atomic Cluster Expansion (ACE) models, which are sampled from the Bayesian posterior on the ACE weights. They use this committee to specify a biasing potential derived from the committee variance, which then allows for acquisition of structures with large predicted error through biased Molecular Dynamics (MD) simulations. Structures are sampled from MD by means of a selection function with some initialised triggering tolerance. The main drawback of this approach is that it is a ``one-shot" process - the full committee is be retrained based on the DFT results of the selected structure after each selection, which makes the scalability of such a solution limited.

To investigate the effectiveness of sampling techniques on this problem, we propose several methods which combine an atomistic descriptor with a standard sampling technique to provide a training dataset. Each method was applied to structures targeting surfaces in the 2018 Silicon GAP dataset \cite{SiGAP} in order to generate a sub-dataset. Each sub-dataset was combined with bulk structures from the same original dataset to form a complete training dataset for an MLIP model. 

Models trained on each of the datasets were then benchmarked based on the RMSE errors on energies and forces applied to all bulk and surface structures, as well as the predicted (100) and (111) surface energies.

\section{Sampling Methodology}
Although the Si dataset contains energy and force information for each structure (and a subset contain virial stresses), these properties are not used to inform sampling methods applied here. The aim of the work was to evaluate methods which only rely on atomic positions, such tha we could perform the selection prior to DFT calculations. Each sampling method uses a local, atom-centered descriptor to define a structure-level feature vector using the average of the atomic features. 

Table \ref{tbl:desc} gives an overview of the descriptors used to generate structure-level feature vectors, which includes the length of the feature vector generated, as well as whether the representation is ``learnable" (i.e. is allowed to adapt during model training to best fit the data)

ACE \cite{ACE} is a fixed-form descriptor used to construct ACE linear models. We use three different parameterisations of the descriptor to test the tradeoff between increased information contained in the descriptor vector, and the corresponding increase in dimensionality. 

\begin{table}[]
\caption{Parameterisation of ACE descriptors, and lengths of all descriptor vectors}
\begin{tabular}{lcccc}
Descriptor         & Order             & Degree            & Learnable           & Length \\
\hline
ACE (S): ``Small"  & 3                 & 12                & X                   & 211               \\
ACE (M): ``Medium" & 3                 & 16                & X                   & 668               \\
ACE (L): ``Large"  & 3                 & 19                & X                   & 1429              \\
MACE (C): ``Core"  &                   &                   & \checkmark          & 640               \\
MACE (T): ``Total" &                   &                   & \checkmark          & 640               \\
MP0 Foundation     &                   &                   & \checkmark          & 640              
\end{tabular}
\label{tbl:desc}
\end{table}

MACE \cite{MACE} uses a learnable atomic embedding in a message passing network, with the representation initially based on the ACE descriptor to form the equivalent of a descriptor. The MP0 foundation model \cite{MP0} was used as an initial baseline for the performance of MACE in the sparsification task. Two bespoke MACE models were trained on Si bulk (``Core" dataset), and Si bulk + surfaces (``Total" dataset), in order to evaluate whether using system-specialised descriptors provides any benefit.

Table \ref{tbl:dataset} shows which ``config types" (labelled subsets of the dataset) were used to inform the common ``Core" dataset used by all models, as well as which extra config types were used in the sampling. The ``Total" dataset is the core dataset plus the sampling dataset. The table also gives information about the number of structures in each config type, as well as the number of atoms in each structure of that config type (multiple rows mean the config type contains structures of different sizes).

\begin{table}[]
\caption{Table showing the conversion from the full 2018 database into the ``Core" Dataset, and the ``Total" = Core + ``Sampling" dataset. The Core dataset covers bulk, elasticity, and some low temperature bulk MD data. The Sampling dataset contains information about surfaces.}
\begin{tabular}{lcccc}
2018 Database                       & ``Core"                     & ``Total"                    &               & \# Atoms \\
Config Type                         & Dataset                     & Dataset                     & \# Structures & per Structure \\
\hline
isolated\_atom                      & \checkmark                  & \checkmark                  & 1             & 1   \\
\hline
\multirow{4}{*}{dia (Si Bulk)}      & \multirow{4}{*}{\checkmark} & \multirow{4}{*}{\checkmark} & 104           & 2   \\
                                    &                             &                             & 220           & 16  \\
                                    &                             &                             & 110           & 54  \\
                                    &                             &                             & 55            & 128 \\
\hline
surface\_001                        &                             & \checkmark                  & 29            & 144 \\
\hline
surface\_110                        &                             & \checkmark                  & 26            & 108 \\
\hline
surface\_111                        &                             & \checkmark                  & 47            & 96  \\
\hline
surface\_111\_pandey                &                             & \checkmark                  & 50            & 96  \\
\hline
surface\_111\_3x3\_das              &                             & \checkmark                  & 1             & 52  \\
\hline
\multirow{3}{*}{decohesion}         &                             & \multirow{3}{*}{\checkmark} & 11           & 16   \\
                                    &                             &                             & 11           & 24  \\
                                    &                             &                             & 11           & 32  \\              
\hline
Total (``Core" Dataset)             &                             &                             & 490          &     \\
\hline
Total (``Total" Dataset)            &                             &                             & 676          &     \\
\end{tabular}
\label{tbl:dataset}
\end{table}

From the set of structure level features, common sparsification techniques were applied in order to provide sampling of the structures. We choose Simple Random Sampling (SRS) as a control, as it is entirely uninformed by the structure features. The $k$-medoids \cite{kmedoids} algorithm partitions the set of features into $k$ clusters (where in this case $k$ is the number of structures we wish to select), and returns the median coordinate of each cluster. Farthest Point Sampling (FPS) \cite{FPS} requires some initial state, often by choosing a single feature vector randomly - in this case we initialise with the set of core dataset features. For each iteration, the method defines a distance metric between each candidate feature and the current state, and selects the candidate feature with the largest associated distance.

We also test an approach derived from the CUR method by Mahoney and Drineas \cite{CUR}. They use a normalised statistical leverage score as a probability distribution, with which they can sample columns and rows of a matrix $A$ in order to produce an accurate CUR decomposition. We can use this scoring scheme to provide a probability distribution over features, and then sample from this distribution in order to select structures. We test two variants of this approach: the first is to use the structure level features similarly to the preceding approaches, and the second is to calculate scores on the atomic descriptor features directly, and sum scores for each feature in a structure.

\subsection{HAL-style Bayesian selection}
We can also define another sampling strategy using ideas from Bayesian linear regression \cite{BishopBLR}, and the posterior covariance. Given a linear model in the total energy $y = E(x) = \sum_{ij} \Phi(x)_{ij} \alpha_j$, with some design matrix $\Phi(x)_{ij} = \phi_j(x_i)$ formed of basis functions $\{\phi_j(x)\}$ of the atomic descriptor vector $x$, and model weights $\alpha$, we can analytically compute the Bayesian posterior distribution $\alpha \sim \mathcal{N}(\mu, \Sigma)$, where:

\begin{equation}
    \Sigma^{-1} = \Phi^T \Lambda \Phi + \Sigma_0^{-1}
\end{equation}
\begin{equation}
    \mu = \Sigma \Phi^T \Lambda^{\frac{1}{2}} y
\end{equation}
From this, it is apparent that the posterior covariance $\Sigma$ does not depend on the energy observations $y$ - only the design matrix $\Phi$ (which is formed of descriptor vectors, e.g. using the ACE or MACE descriptors from Table \ref{tbl:desc}), prior covariance matrix $\Sigma_0$, and likelihood precision matrix $\Lambda$ are required (the prior covariance and likelihood precision can be expressed as hyperparameters of the Bayesian model).

For some new candidate observation $(x^*, y^*)$, we can define a score function based on the design matrix $\Phi^*$ associated with this new observation. Evaluating this score function on a set of candidate structures, we select the structure with the highest associated score, and update the posterior covariance
\begin{equation}
    \label{eqn:post_update}
    \Sigma_{n+1}^{-1} =
    \begin{pmatrix}\Phi_n \\ \Phi^*\end{pmatrix}^T
    \begin{pmatrix} \Lambda_n & 0 \\ 0 & \Lambda^*\end{pmatrix} 
    \begin{pmatrix}\Phi_n \\ \Phi^*\end{pmatrix} + \Sigma_0^{-1}
    \equiv \Phi^{*T} \Lambda^* \Phi^* + \Sigma_{n}^{-1}
\end{equation}

Here, $\Sigma_{n}$ is the posterior covariance matrix of the $n^\text{th}$ iteration, trained with design matrix $\Phi_n$, and $\Phi^*$ is the design vector of the newly selected point $x^*$

We can use the posterior in a manner similar to HAL, by defining the score function $s$ based on the posterior predictive uncertainty of observables. Here we use the variance in per-atom energy 
\begin{equation}
    \label{eqn:peratom}
    s = \mathrm{Var}(E_\text{per-atom}) = \frac{1}{N^2} \sum_{ij} \left(\Phi^* \Sigma \Phi^{*T}\right)_{ij}
\end{equation}

\section{Results}
\begin{figure}[htb]
    \centering
    \includegraphics[width=1\linewidth]{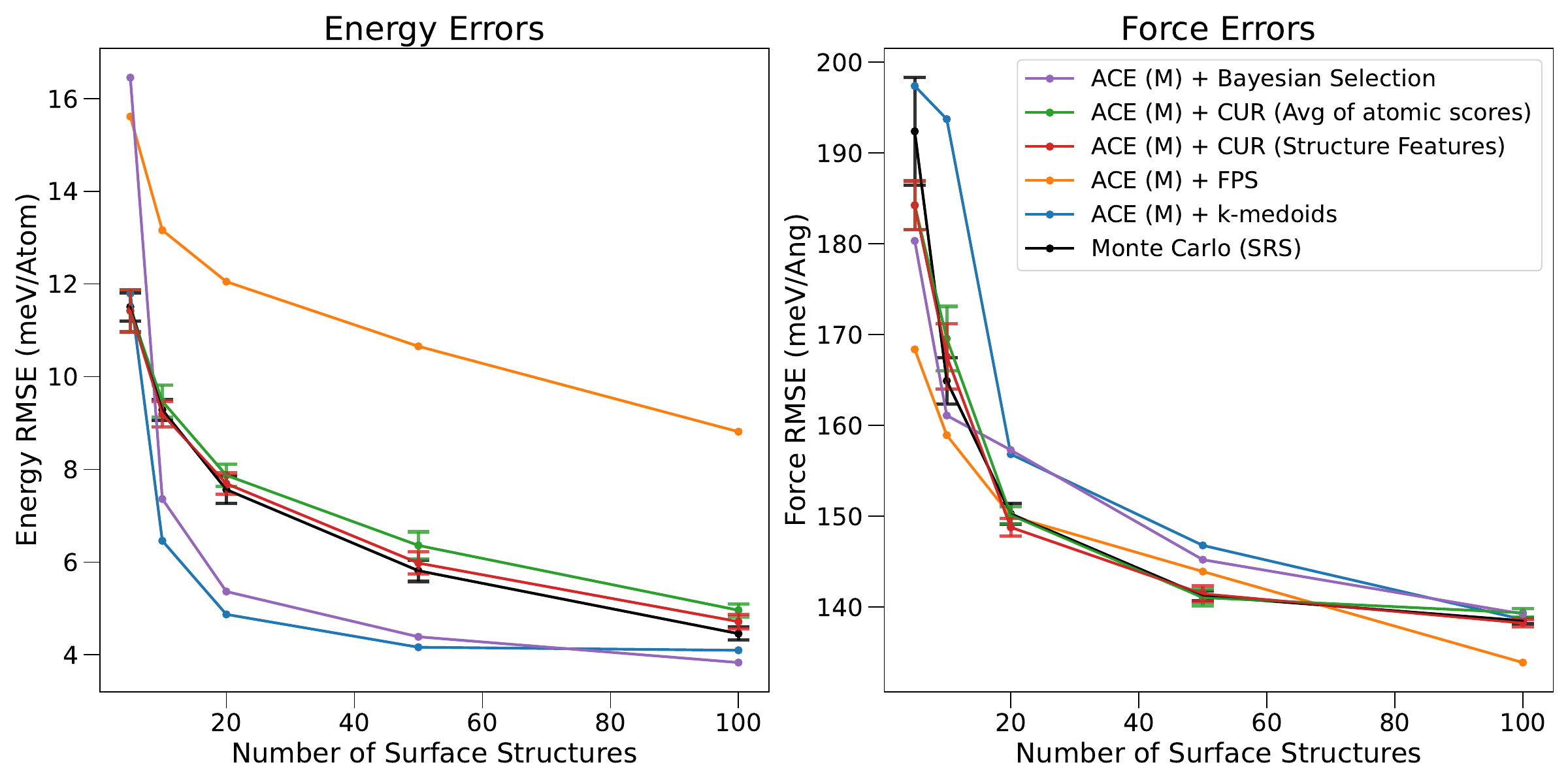}
    \caption{A comparison of the energy (left) and force (right) RMSE errors across the ``Total" dataset, using the ``Medium" parameterisation of the ACE descriptor as an input space for various standard sparsification methods.}
    \label{fig:ErrsMethod}
\end{figure}

\begin{figure}[htb]
    \centering
    \includegraphics[width=1\linewidth]{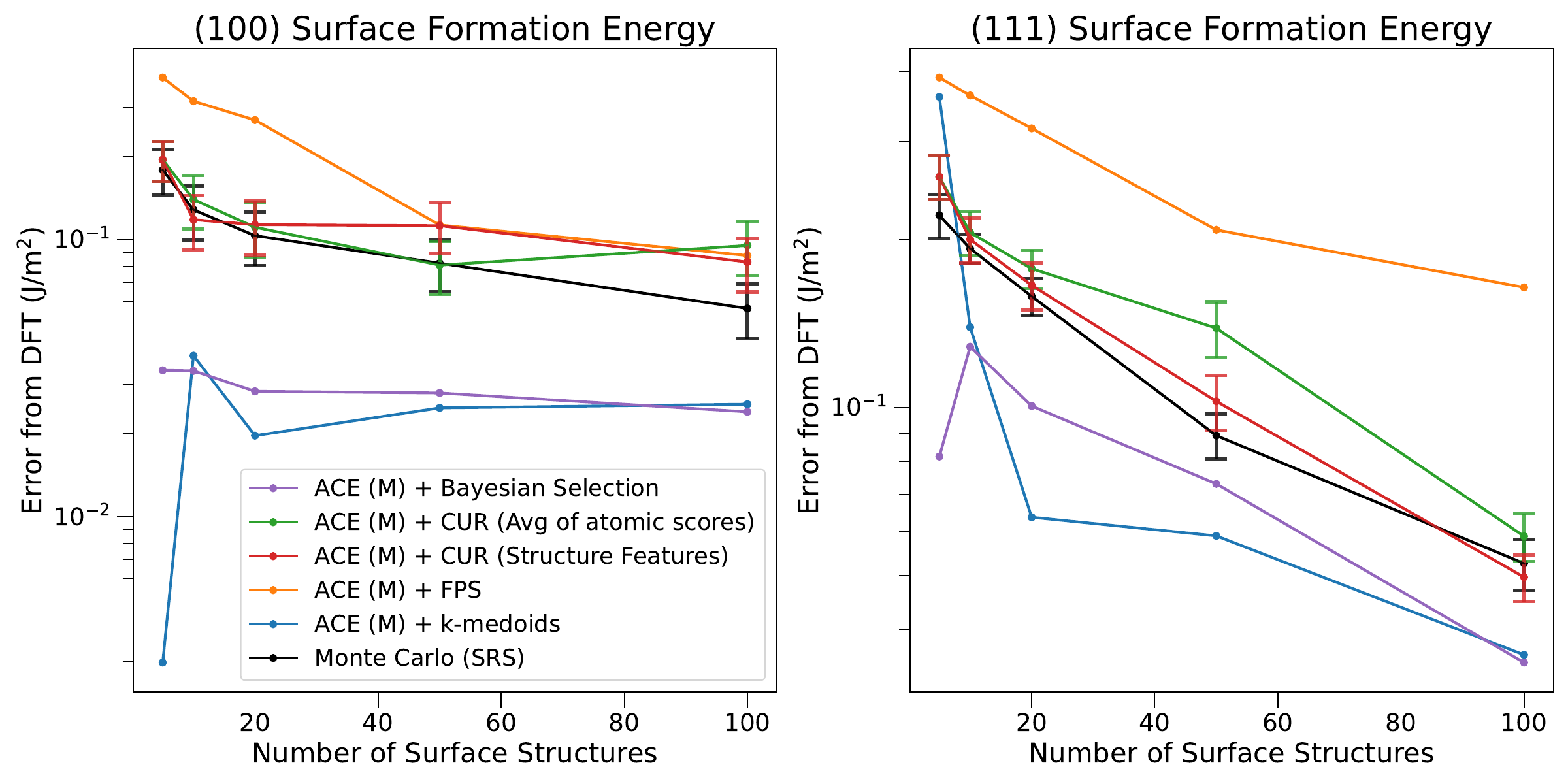}
    \caption{A comparison of the error on predictions of the (100) (left) and (111) (right) surface energies, using the ``Medium" parameterisation of the ACE descriptor as an input space for various standard sparsification methods.}
    \label{fig:SurfMethod}
\end{figure}

Figures \ref{fig:ErrsMethod} and \ref{fig:SurfMethod} compare the performance of using the ACE descriptor as an input to each of the sampling methods, at a range of different dataset sizes (each dataset is the full ``Core" dataset, plus a small number of surface structures sampled from a larger pool of available data). 

Figure \ref{fig:ErrsMethod} makes this comparison in terms of the RMSE error on energies and forces across all of the ``Core" bulk structures, and the full pool of surface structures. The sampling methods have nearly equivalent force errors, but Bayesian selection is approximately 5~meV/\AA{} worse across the dataset for $N=20$, where $N$ is the number of surface structures included in training, and $k$-medoids being around 30~meV/\AA{} worse at $N=10$ (both methods get much closer to average when the dataset size is increased). The energy errors show a different picture, with $k$-medoids and Bayesian selection being better for $N\geq5$ (k-medoids is around 2.5~meV/atom better than SRS at $N=20$). Both CUR-based methods perform comparably with SRS across all dataset sizes, and FPS sampling performs consistently poorly in terms of energy errors when compared to SRS ($\approx$4.8~mev/atom worse at $N=20$; similar difference at $N=100$).

Figure \ref{fig:SurfMethod} compares the error on predicted (100) and (111) surface energies when compared to DFT, for the same set of sampling methods using the ACE descriptor. Here, $k$-medoids and Bayesian selection do very well at predicting the (100) surface energy, consistently giving results to an error of under 0.05~J/m$^2$ for all $N$. The methods are also the best performing for the (111) surface energy, with $k$-medoids incurring less error for $N>10$, but having significantly higher error for $N=5$. Again CUR performs comparably to SRS, with predictions of the (100) surface being on average worse. Using CUR with structure features appears to work slightly better than SRS on the (111) surface, but the two results are extremely close. FPS consistently performs the worst on the (111) surface, with an error at $N=100$ that is around 0.18~J/m$^2$ higher than CUR with score averaging, which is the next worst performing method here. FPS is also poor at predicting an accurate (100) surface energy for $N<50$, after which it constructs datasets of similar quality to the CUR approaches.

\begin{figure}[htb]
    \centering
    \begin{subfigure}{.98\linewidth}
        \includegraphics[width=\linewidth]{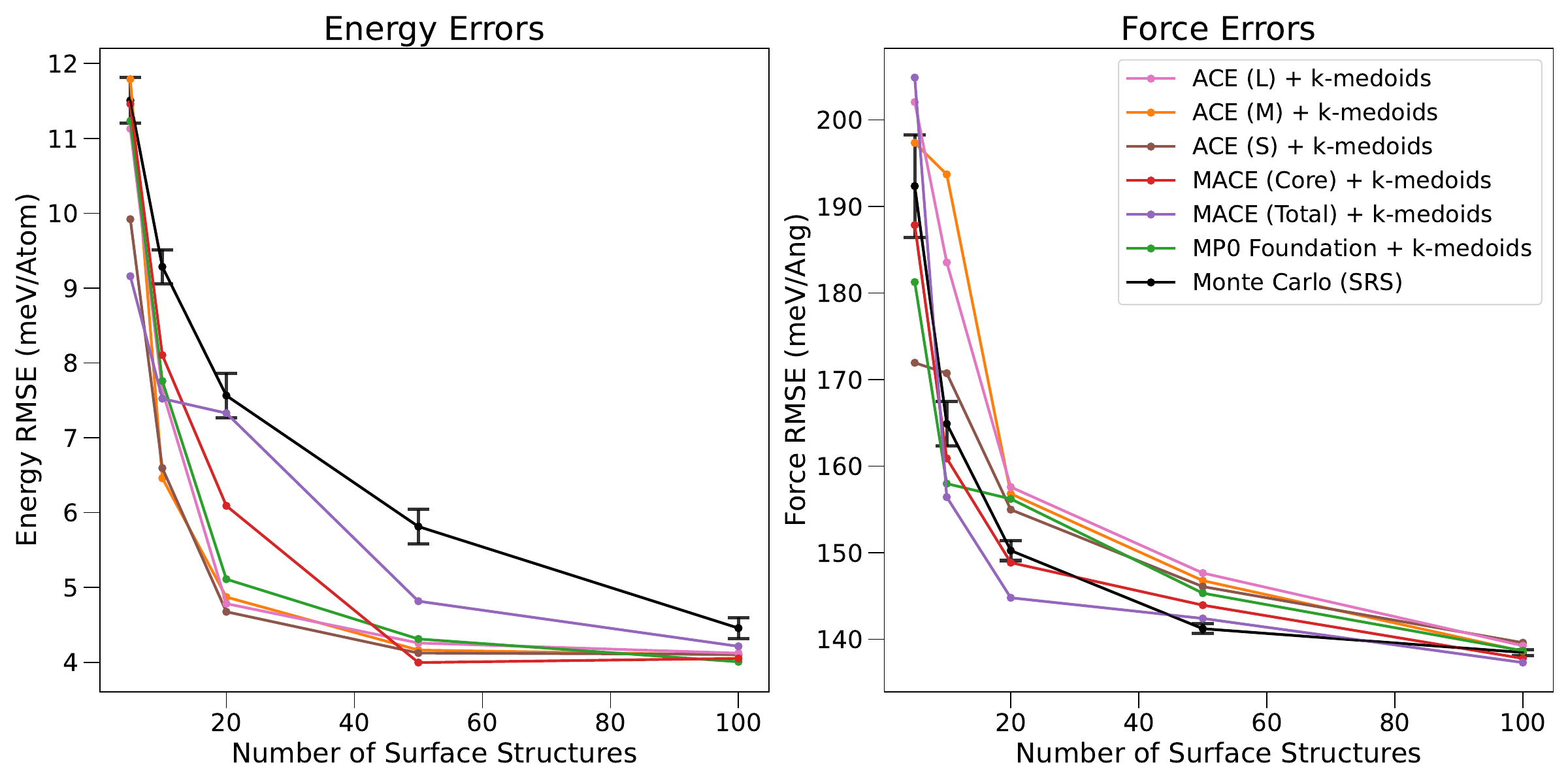}
        \caption{Dataset RMSE Errors, using $k$-medoids sampling on different descriptors}
        \label{fig:ErrsDescKMED}
    \end{subfigure}
    \hfill
    \begin{subfigure}{.98\linewidth}
        \includegraphics[width=\linewidth]{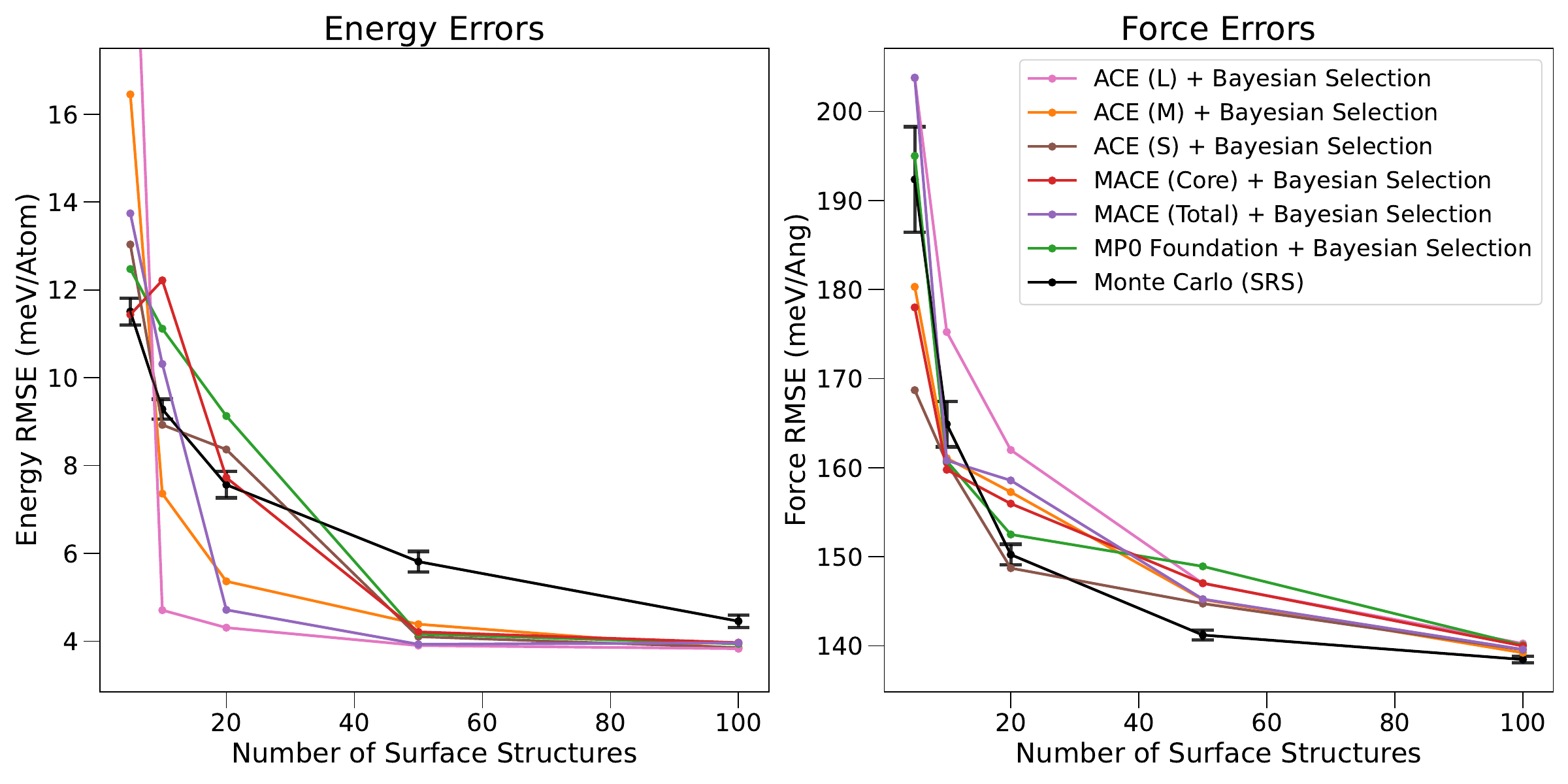}
        \caption{Dataset RMSE Errors, using Bayesian selection on different descriptors}
        \label{fig:ErrsDescBLR}
    \end{subfigure}
    \caption{Comparison of the energy and force RMSE errors over the ``Total" dataset. Each subplot shows a range of common descriptor functions used to provide an input to each of the standard sparsification methods.}
    \label{fig:ErrsDesc}
\end{figure}

To compare the relative performance of each descriptor, we limit the methods to $k$-medoids and Bayesian selection as these appear to be the best performing methods in the ``Medium" ACE descriptor. Figure \ref{fig:ErrsDescKMED} shows the overall energy and force RMSE errors for each of the choices of descriptor applied to the $k$-medoids method. We see that all of the descriptors perform better than SRS in terms of energy RMSE, and near equivalent in terms of force RMSE. Most descriptors perform very similarly, but the ``Total" MACE descriptor appears to be the worst performing at predicting energies, but the best at predicting forces.

Figure \ref{fig:ErrsDescBLR} shows the equivalent comparison for the Bayesian selection method. We see a similar picture for the force error, where the descriptors are approximately equal with SRS, but we see a different picture for energy errors. At $N=5$ we see that the ``Medium" and ``Large" ACE descriptors perform considerably worse than SRS, but that they become significantly better for $N=10$, where the ``Large" descriptor has 4~meV/Atom lower energy RMSE than SRS. 

\begin{figure}[htb]
    \centering
    \begin{subfigure}{.98\linewidth}
        \includegraphics[width=\linewidth]{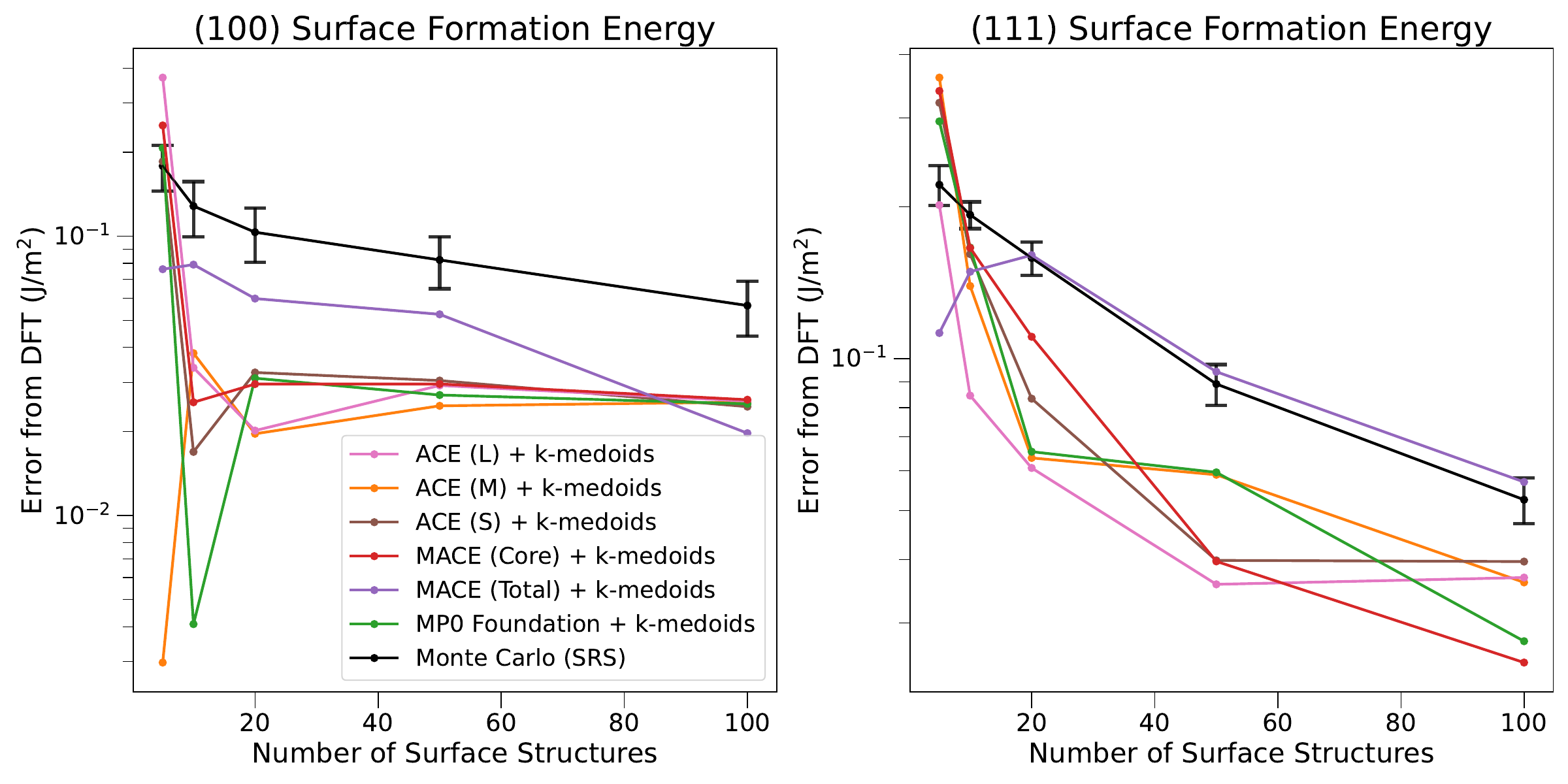}
        \caption{Error on surface energy predictions, using $k$-medoids sampling on different descriptors}
        \label{fig:SurfDescKMED}
    \end{subfigure}
    \begin{subfigure}{.98\linewidth}
        \includegraphics[width=\linewidth]{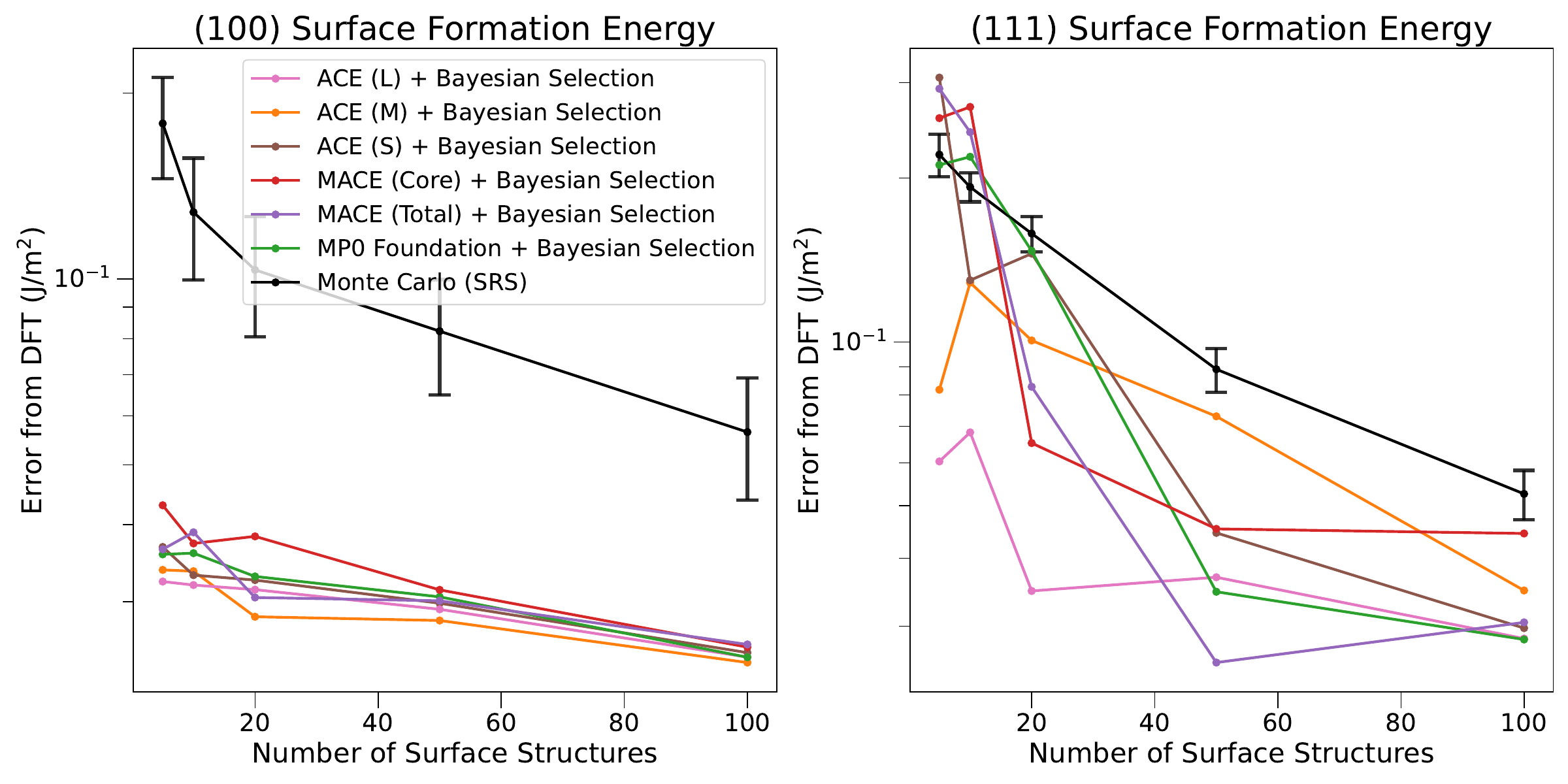}
        \caption{Error on surface energy predictions, using Bayesian selection on different descriptors}
        \label{fig:SurfDescBLR}
    \end{subfigure}
    \caption{Errors on (100) and (111) surface energies (compared with DFT). Each panel shows a range of common descriptor functions used to provide an input to each of the sparsification methods.}
    \label{fig:SurfDesc}
\end{figure}

Figure \ref{fig:SurfDescKMED} compares the descriptors across surface energy predictions. We see that initially the MP0, ``Core" MACE, and ``Large" ACE descriptors perform worse than SRS at predicting the (100) energy for $N=5$, but that all descriptors become much better than SRS for $N>5$. The ``Total" MACE descriptor is the worst performing at predcting the (100) energy for $10 \leq N \leq 50$. For the (111) surface, we see that MP0, ``Core" MACE, and the ``Small" and ``Medium" ACE descriptors perform worse than SRS initially, but almost al of the descriptors become much better than SRS for $N>10$. The exception is the ``Total" MACE descriptor, which performs equivalently to SRS for $N\geq 20$.

Figure \ref{fig:SurfDescBLR} shows the surface energy comparison for the Bayesian selection method. Here, we see that all descriptors perform extremely well at predicting the (100) surface, even for small $N$, and the choice of descriptor does not appear to significantly affect the error. On the (111) surface, we see that all descriptors using BLR perform better than SRS for $N\geq 20$, with the ``Medium" and ``Large" ACE descriptors consistently performing better for all $N$. The ``Core" and ``Total" MACE descriptors also perform competitively for $N \geq 20$, but perform worse than SRS at $N=5$.

\begin{figure}
    \centering
    \includegraphics[width=\linewidth]{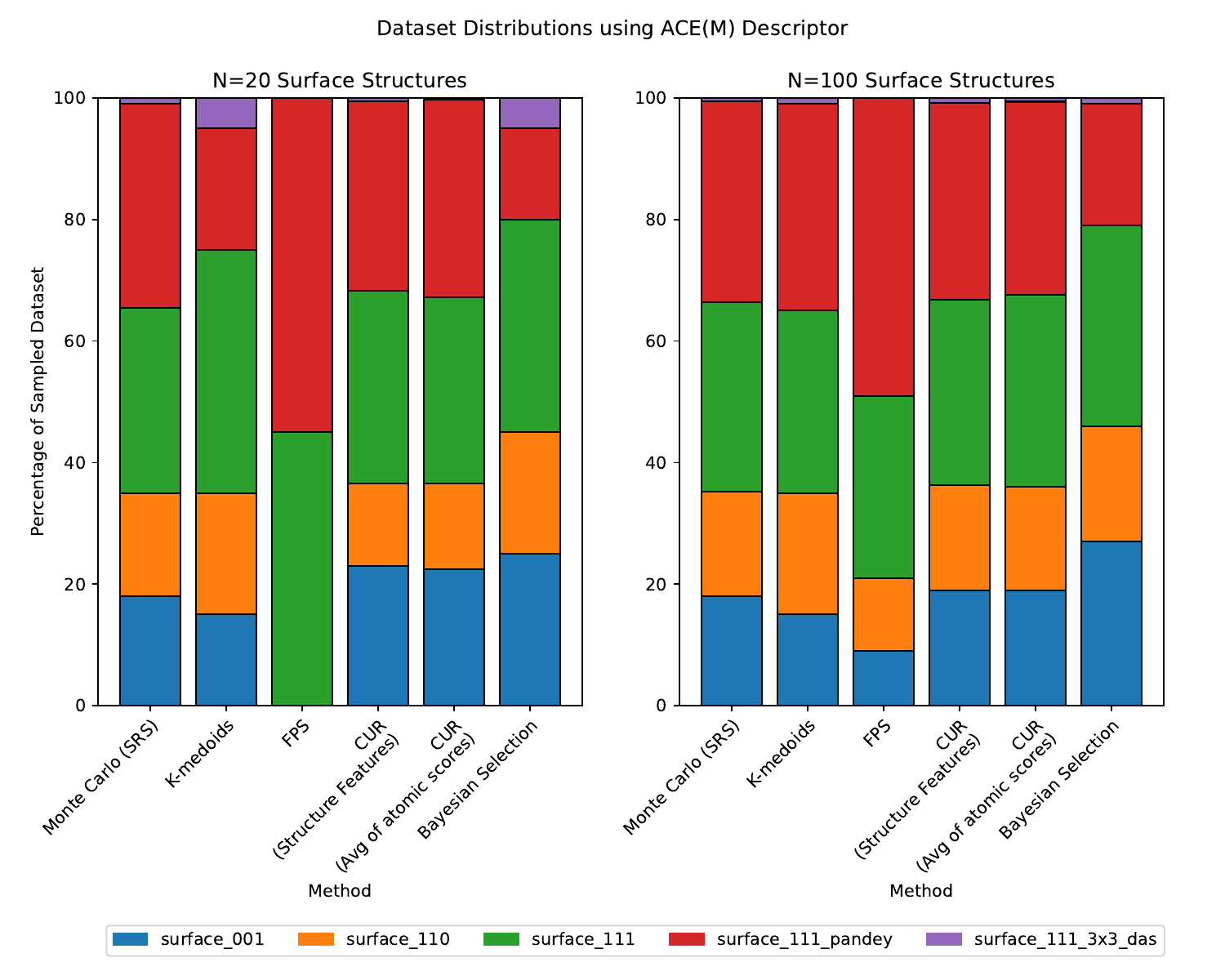}
    \caption{Distributions of configuration types obtained with each sampling method for target dataset size of $N$=20 (left panel) and $N$=100 (right panel). The configuration type labels shown in the legend match those in Table~\ref{tbl:dataset}.}
    \label{fig:datasetDist}
\end{figure}

Figure \ref{fig:datasetDist} shows a breakdown of the sampled datasets for two choices of the number of samples, $N=20$ and $N=100$. For each $N$, we plot the number of structures belonging to each configuration type (as listed in Table~\ref{tbl:dataset}). We see that at $N=20$, the FPS approach exclusively sampled structures targeting reconstructions of the (111) surface, and did not sample anything targeting other surfaces. We see that Bayesian selection sampled the most (001) structures, which could explain the observed high accuracy in Figure \ref{fig:SurfMethod}, but the number sampled is not considerably higher than the CUR approaches. For $N=100$, each of the methods sampled each of the configuration types in reasonable proportions. The fact that, aside from FPS, the methods sample each configuration type in reasonably similar proportions implies that the differences in accuracy observed in the formation energy benchmark depend on aspects more subtle than just the configuration type alone.

\begin{figure}
    \centering
    \includegraphics[width=\linewidth]{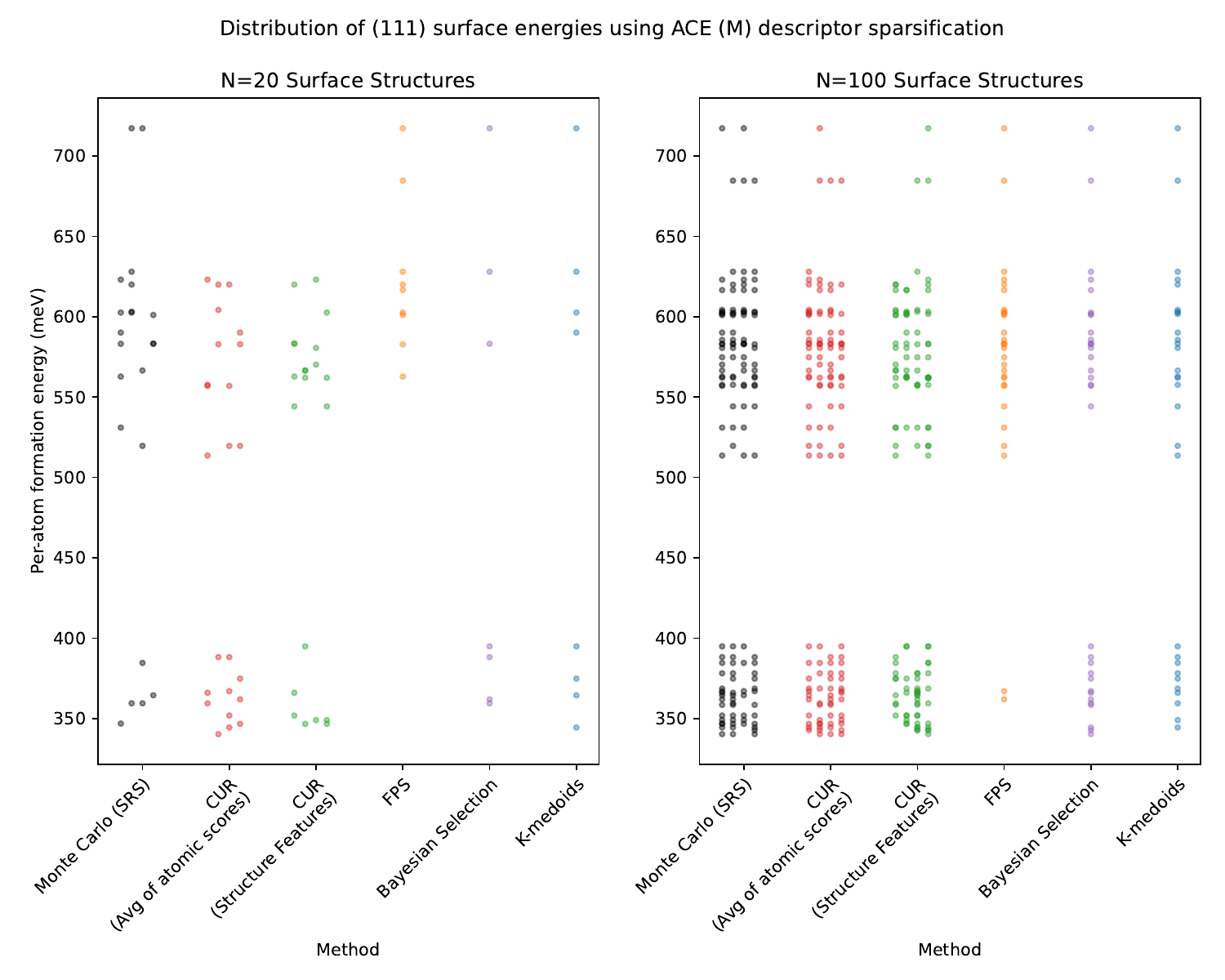}
    \caption{Distribution of the formation energy per atom for (111) surface structures (from the \texttt{surface\_111} configuration type), as selected by each sampling method for target dataset size of $N=20$ (left panel) and $N=100$ (right panel).}
    \label{fig:datasetEnergyDist}
\end{figure}

Figure \ref{fig:datasetEnergyDist} shows a more specialised breakdown, looking only at the \texttt{surface\_111} configuration type. The surface formation energies of the structures selected by each method are shown, again for $N=20$ and $N=100$. As the SRS and CUR approaches involve randomness in the sampling, four representative sampled datasets are shown. It is clear from the plot that the FPS approach is extremely biased against the lower energy structures, which explains the inaccuracies at reproducing an energy minimisation. Whilst the SRS and CUR approaches are able to effectively sample the full energy ranges here, at $N=20$ the extrema are missed, and the resulting dataset is very dependent on the specific random realisation. The K-medoids and Bayesian selection methods select across the full energy range even at $N=20$. 

\section{Discussion}
It is clear from Figs. \ref{fig:ErrsMethod} \& \ref{fig:SurfMethod} that the choice of sparsification method does impact the performance of resulting MLIP models, both in terms of training RMSE errors, and in predictions of key quantities. It is ultimately difficult to argue that any one method is universally the ``best", due to the work focusing on a single test case, but our results support the use of $k$-medoids and Bayesian selection. The results also show that these methods were consistently competitive against SRS across a range of different descriptors.

Although $k$-medoids and Bayesian selection perform comparably on the chosen tests, Bayesian selection may be a better choice for expanding on existing datasets, as we are able to use the posterior formed from the existing dataset as the new prior for the selection process. We could also improve the method by attempting to encode more information into the sparsification process: by including force and/or stress contributions to the Bayesian posterior, and also by customising the scoring function based on desired outcomes (i.e. by using a scoring metric based on forces to target improvements to force predictions, or by using a weighted sum of several metrics).

We also see that the Bayesian selection method generates models which are significantly closer to DFT for the (100) surface energy, when compared to the (111) surface energy. This is likely due to the many physical (111) surface reconstructions, of which three are included in the dataset, and only one is tested.

\subsection{Future Outlook}
There are many ways to adapt the Bayesian selection approach, which may yield some improvement relative to K-Medoids. Firstly, we could try to define a ``better" structure-level descriptor, i.e. one which more effectively partitions different kinds of structures, and/or provides an adequate partitioning in a lower dimensional space. One approach to achieving this would be to use a learnable descriptor, such as a deep kernel model \cite{DeepKernel}. The downside of this approach is that in order for the descriptor to be application-specialised, it essentially would need a training dataset containing exactly the kinds of structure the model would be trained to find. This would then require an iterative approach to dataset development, slowing down the time-to-science.

The other main way to modify the selection approach is to use other score functions - the variance in per-atom energy metric used in this work is competitive with K-Medoids on this dataset, but may not be the optimal choice for other applications. 

Taking inspiration from the work of Subramanyam and Perez, we could use the informational entropy as a selection metric. The entropy of a set of random variables distributed as a multivariate normal is proportional to the log determinant of the covariance matrix, so the corresponding score function would be 
\begin{equation}
    \begin{split}
    s = H_n(\Phi^*_k) &= \log |\Sigma_k^*| \\
    &\equiv \log \left| \left[
    \begin{pmatrix}\Phi_n \\ \Phi^*_k\end{pmatrix}^T
    \begin{pmatrix} \Lambda_n & 0 \\ 0 & \Lambda^*_k\end{pmatrix} 
    \begin{pmatrix}\Phi_n \\ \Phi^*_k\end{pmatrix} + \Sigma_0^{-1}\right]^{-1} \right| \\
    &= - \log \left| \Sigma_n^{-1} +  \Phi^{*T}_k \Lambda^*_k \Phi^*_k\right|
    \end{split}
\end{equation}

Alternatively, we could define score functions based on other observables, similar to the per-atom energy variance used above. One strong candidate metric would be the maximum force variance
\begin{equation}
    \label{eqn:maxforce}
    s = \max_i(\mathrm{Var}(F_i)) = \max_i \left[ \left(\Phi_F^* \Sigma_n \Phi_F^{*T}\right)_{ii} \right]
\end{equation}
where $\Phi_F$ is the design matrix of the force features (which are sums over derivatives of the descriptor vector $\frac{\partial (\Phi_{E})_i}{\partial r_j}$).

If we have an existing model for the system, e.g. a foundation model, classical potential, or a previous iteration of a bespoke model (i.e. some means of estimating the energies and forces with reasonable accuracy), we can directly use the same selection metric as HAL
\begin{equation}
    s = f_i = \frac{(\Phi_F^* \Sigma \Phi_F^{*T})_{ii}}{||\bar{F_i}|| + \epsilon}
\end{equation}
as $\Phi_F^* \Sigma \Phi_F$ is equivalent to the $\frac{1}{k}\sum_k||F_i^k - \bar{F_i}||$ variance over the HAL committee. Here, $\bar{F_i}$ is the mean force observation (i.e. the prediction of the existing foundation, classical, or bespoke model), and $\epsilon$ is a small constant which prevents divergence for $||\bar{F_i}|| \rightarrow 0$.

We can use this approach to generalise the ACEHAL approach: given any general MLIP model $E_\text{MLIP}(x) = f_\text{MLIP}(x,\alpha_\text{MLIP})$ ($f_\text{MLIP}$ could be nonlinear in the weights), we can define a linear surrogate in terms of the atomic descriptor $x$, with posterior covariance
\begin{equation}
    \begin{split}
    &\Sigma^{-1} = \Phi^T \Lambda \Phi + \Sigma_0^{-1}
    \end{split}
\end{equation}
where $\Phi_{ij} = \{x_j\}_i$ is a design matrix formed of the atomic descriptor evaluations.

We could then perform biased MD simulations exactly as HAL does: using the true MLIP model to describe the true potential energy surface, and using the linear surrogate to inform the uncertainty-based biasing potential. We can also update the posterior covariance (using Eqn. \ref{eqn:post_update}) of the surrogate with each structure selected. 

One key advantage of this procedure is the relaxation of the HAL ``one-shot" scheme to an ``$N$-shot" scheme (where $N$ is the number of structures added to the dataset before the MLIP model is retrained), increasing the throughput of the process by deferring DFT and refits to occur in batches, instead of after every selection. 

With standard HAL, if we had instead selected $N$ structures without refitting the committee, we would not account for correlations in the structures selected. We therefore may oversample regions of previously high uncertainty, which may have only required a single structure to correct. The Bayesian selection approach fixes this oversampling issue by providing an update to the posterior covariance (which is what defines the variance of the committee and thus the sampling metric) after each of the structure is drawn. 

Since we build a linear surrogate to form the biasing potential, we are not restricted to descriptor features from linear models - we are also able to leverage learnable descriptor features such as the features of the MACE descriptor.

\section{Conclusions}
A new Bayesian selection method, derived from Bayesian linear regression, was developed for the purpose of sampling structures from a candidate set, with the aim of producing a concise dataset which maximally improves MLIP model accuracy. We show that the method is effective at producing ``good" datasets, where models trained on these datasets accurately reproduce targeted quantities of interest, for the given test case. We also show that the method performs better than some competing approaches at the given task, and that it remains similarly effective for a number of atomic descriptors.

\section{Data Availability}
Data and scripts required to reproduce this work are provided by a GitHub repository, and corresponding Zenodo archive \cite{Zenodo}.

\section{Acknowledgments}
T.R is supported by a studentship within the Engineering and Physical Sciences Research Council supported Centre for Doctoral Training in Modelling of Heterogeneous Systems, Grant No. EP/S022848/1, with additional funding from Huawei Technologies R\&D UK.

We acknowledge usage of the ARCHER2 facility for which access was obtained via the UKCP consortium and funded by EPSRC Grants No. EP/P022561/1 and No. EP/X035891/1. Calculations were performed using the Sulis Tier 2 HPC platform hosted by the Scientific Computing Research Technology Platform at the University of Warwick. Sulis is funded by EPSRC Grant EP/T022108/1 and the HPC Midlands+ consortium. Further computing facilities were provided by the Scientific Computing Research Technology Platform of the University of Warwick.

\section{References}
\printbibliography[heading=none]
\end{document}